\documentclass[a4paper,11pt]{article}
\usepackage{jcappub}
\usepackage{graphicx}
\usepackage{bbm}
\usepackage{bm}
\usepackage{amsfonts}
\usepackage{epstopdf}
\usepackage{mciteplus}
\usepackage{hyperref}
\usepackage{enumerate}
\newcommand\diff{\mathrm{d}}

\makeatletter
\def\@fpheader{\relax}
\makeatother

\title{
	Spherical collapse and virialization in $f(T)$ gravities
}
\author[a]{Rui-Hui Lin,}
\author[a]{Xiang-Hua Zhai,}
\author[a]{Xin-Zhou Li}
\affiliation[a]{Shanghai United Center for Astrophysics (SUCA), Shanghai Normal University,
100 Guilin Road, Shanghai 200234, China}

\emailAdd{1000379711@smail.shnu.edu.cn}
\emailAdd{zhaixh@shnu.edu.cn}
\emailAdd{kychz@shnu.edu.cn}

\abstract{
Using the classical top-hat profile,
we study the non-linear growth of spherically symmetric density perturbation
and structure formation in $f(T)$ gravities.
In particular, three concrete models,
which have been tested against the observation
of large-scale evolution and linear perturbation of the universe
in the cosmological scenario,
are investigated in this framework,
covering both minimal and nonminimal coupling cases of $f(T)$ gravities.
Moreover,
we consider the virialization of the overdense region in the models
after they detach from the background expanding universe
and turn around to collapse.
We find that there are constraints
in the magnitude and occurring epoch
of the initial perturbation.
The existence of these constraints indicates that
a perturbation that is too weak or occurs too late
will not be able to stop the expanding of the overdense region.
The illustration of the evolution of the perturbation
shows that in $f(T)$ gravities,
the initial perturbation within the constraints
can eventually lead to clustering and form structure.
The evolution also shows that
nonminimal coupling models collapse slower than
the minimal coupling one.
}
\begin{document}

\keywords{$f(T)$ theory; top-hat collapse; virialization}
\maketitle
\section{Introduction}
At the end of the last century,
the astronomical observation of
high redshift type Ia supernovae (SNeIa)
indicated that our universe is not only expanding,
but also accelerating,
which conflicts with our deepest intuition of gravity.
Current observations,
such as cosmic microwave background (CMB),
baryon acoustic oscillations (BAO)
and large scale structure (LSS),
converge on the fact of accelerating expansion.
In order to make reasonable sense of this acceleration,
an exotic component of the universe,
the mysterious dark energy (DE),
has been introduced.
Although many efforts have been made,
the identity and physical nature of DE
still seem to evade disclosure.
Besides seeing DE as a real content of our universe,
one possible way to address the problem
is to modify gravitation theory
such that the acceleration could be attributed to this modification
(for review, see, e.g., \cite{Capozziello2011}).

One of the extensively studied modified gravities is the $f(R)$ gravity,
in which one starts from the standard General relativity (GR),
and extends the standard Hilbert-Einstein action to
an arbitrary function of the Ricci scalar $R$\cite{Sotiriou2010,DeFelice2010,Nojiri2011}.
A further extension of $f(R)$ gravity proposed in \cite{Nojiri2004,Allemandi2005} and then studied in \cite{Bertolami2007,Harko2008,Harko2010,Bertolami2010} is to consider the nonminimal coupling
between matter and gravity,
i.e. the $f(R,\mathcal{L}_\text{m})$ gravity,
where $\mathcal{L}_\text{m}$ is the Lagrangian density of matter.
On the other hand,
one can also modify the gravitation theory
starting from the Teleparallel Equivalent of General Relativity (TEGR)
\cite{Einstein1928,Hayashi1979,aldrovandi2012teleparallel,Maluf2013}.
In the Lagrangian of TEGR,
the torsion scalar $T$ takes the place of the Ricci scalar $R$,
and hence in analogous to $f(R)$,
$f(T)$\cite{Bengochea2009,Linder2010,Cai2016}
and nonminimal coupling $f(T,\mathcal{L}_\text{m})$ gravities
\cite{Harko2014a,Feng2015}
have been studied.
Despite the equivalence between TEGR and GR,
$f(T)$ gravities are different from $f(R)$ gravities.
One of the important advantages of $f(T)$ gravities
is that the field equations are second order instead of fourth order.
It is also noticed that
non-trivial $f(T)$ gravities violate
the Lorentz invariance\cite{Li2011,Sotiriou2011,Bahamonde2015} and
particular choices of tetrad are important
to get viable models\cite{Tamanini2012}.

Concerning large scale evolution of the universe
and the linear perturbations of the background,
$f(T)$ gravities have been compared with the cosmological observation
data including CMB, BAO and SNeIa
\cite{Bengochea2009,Feng2015,Nunes2016}.
While the SNeIa data capture the late time accelerating behavior
(redshift $z<3$),
and the data of BAO and inhomogeneity of CMB
provide mostly the imprints of
the early development and linear perturbations of the background
before recombination ($z>1000\sim1500$),
most structures such as stars, galaxies and clusters of galaxies
are formed from the non-linear evolution of perturbations
during the post-recombination epoch,
which is usually referred as
the Dark Ages ($10<z<1000$).
However, a fully relativistic treat of this
non-linear perturbation is not currently available.
Thus it is usually handled by $N$-body simulation
(see e.g. \cite{Maccio2004,Aghanim2009,Baldi2010,Li2011a}),
which can be cumbersome and time-consuming,
and is not practical for one to use to study different gravitation models.
On the other hand,
there has been intermediate approximations to
the full-fledged theory before turning to simulation,
one of which is spherical collapse model
\cite{Gunn1972,Padmanabhan1993}.
In this simple semi-analytic model,
a spherical overdense region evolves with the expanding background universe,
and slows down and turns from expanding to collapsing.
As a simple but fundamental tool to describe
the growth of gravitationally bounded systems,
this model has been used in various studies of gravitation theories
in recent years
\cite{Mota2004,Maor2005,Nunes2006,Basilakos2009,Basilakos2010,Brax2010,Fernandes2012,Bellini2012,Li2014,Carames2014}.
The model has also been compared with the pseudo-Newtonian approach\cite{Abramo2007,Abramo2009}
and it is found that the two approximation schemes convey
identical equations for the density contrast.

After the turning around from expanding to collapsing,
the overdense region shall virialize and be prevented from
falling into a singularity.
For gravitationally bounded systems,
virialization is also a powerful approach
with a long history.
In fact,
the discrepancy between the virial mass $M_\text{vir}$
and the total baryonic mass $M_\text{b}$ of a cluster
contributed to the awareness of the existence of dark matter.
However, the virial theorem depends on the gravitation theory,
and hence the corresponding theorems in different modified gravities
have drawn a lot of attention lately
\cite{Bohmer2008,Sefiedgar2009,Capozziello2013,Milgrom2014,Santos2015,Javadinezhad2016}.
In the context of spherical collapse,
virialization provides a more realistic picture
of the collapsing structures,
and enable ones to calculate the main features of them
(collapse factor, virial density, virial mass, etc.),
which may be useful in further analytic investigation
or numerical simulation
(e.g. NFW profile\cite{Navarro1997} to describe the $N$-body simulation).

In this paper,
we look into the collapse of a spherically symmetric perturbation
for models in $f(T)$ gravities
with a classical top-hat profile,
including both minimal and nonminimal matter-gravity coupling cases.
Moreover, we consider the virialization of the collapse
after the turn-around point.
The paper is organized as follows:
in Section \ref{review},
we briefly review the bases and cosmologically tested models
of the $f(T)$ gravities,
including both minimal and nonminimal matter-gravity coupling cases.
In Sec.\ref{collapse} , we consider the spherical top-hat collapse
and evolution of perturbation in these models.
We study the virialization of the collapse in Sec.\ref{virial} .
Sec.\ref{conclusion} contains our conclusions and discussions.

We are going to use the Latin letters
$(a,b,c,\cdots=0,1,2,3)$
to denote the tangent space indices,
and Greek letters
$(\mu,\nu,\rho,\cdots=0,1,2,3)$
to denote the spacetime indices.
We assume the Lorentz metric of Minkowski spacetime
\begin{equation}
	\eta=\eta_{ab}\diff x^a\otimes\diff x^b
	\label{eta}
\end{equation}
has the form
$\eta_{ab}=\text{diag}(+1,-1,-1,-1)$.
And we use the unit
$c=1$
throughout the paper.

\section{From TEGR to $f(T)$ models}
\label{review}
\subsection{The field equations}
In the torsional formulations of gravity,
one uses tetrad fields $\{e_a,e^a\}$
as the fundamental dynamical variables.
The tetrad fields form an orthonormal base
of the tangent space $T_p\mathcal{M}$ at each point $p$
on the spacetime differentiable manifold $\mathcal{M}$.
The spacetime metric $g$
\begin{equation}
	g=g_{\mu\nu}\diff x^\mu\otimes\diff x^\nu
	\label{metricg}
\end{equation}
is related to the tangent space metric $\eta$ by
\begin{equation}
	\eta_{ab}=g(e_a,e_b)=g_{\mu\nu}e_a^{\:\mu}e_b^{\:\nu},
	\label{metrictrans}
\end{equation}
or conversely,
\begin{equation}
	g_{\mu\nu}=\eta_{ab}e^a_{\:\mu}e^b_{\:\nu}.
	\label{metrictrans1}
\end{equation}
And hence the determinant
\begin{equation}
	|e|\equiv\text{det}(e^a_{\:\mu})=\sqrt{-g}.
	\label{metricdet}
\end{equation}
One then can use the tetrad field to define
the curvatureless Weitzenb{\"o}ck connection\cite{Weitzenbock1923}
\begin{equation}
	\tilde{\Gamma}_{\nu\mu}^{\lambda}\equiv e_a^{\:\lambda}\partial_\mu e_{\:\nu}^a=-e_{\:\nu}^a\partial_\mu e_a^{\:\lambda}
	\label{connection}
\end{equation}
instead of the torsionless Levi-Civi{\'a} one.
The torsion and contorsion tensors are then given by
\begin{equation}
	T_{\:\:\mu\nu}^\lambda\equiv\tilde{\Gamma}_{\nu\mu}^\lambda-\tilde{\Gamma}_{\mu\nu}^\lambda=e_i^\lambda(\partial_\mu e_\nu^i-\partial_\nu e_\mu^i),
	\label{torsiontensor}
\end{equation}
and
\begin{equation}
	K^{\mu\nu}_{\quad\rho}\equiv-\frac12\left(T^{\mu\nu}_{\quad\rho}-T^{\nu\mu}_{\quad\rho}-T_{\rho}^{\:\mu\nu}\right),
	\label{contorsion}
\end{equation}
respectively.
Utilizing the tensor
\begin{equation}
	S_\rho^{\:\mu\nu}\equiv\frac12\left(K^{\mu\nu}_{\quad\rho}+\delta_\rho^\mu T^{\lambda\nu}_{\quad\lambda}-\delta_\rho^\nu T^{\lambda\mu}_{\quad\lambda}\right),
	\label{auxtensor}
\end{equation}
one can define the torsion scalar
\begin{equation}
	T\equiv T^\rho_{\:\mu\nu}S_\rho^{\:\mu\nu}.
	\label{torsionscalar}
\end{equation}
TEGR\cite{Hayashi1979,aldrovandi2012teleparallel,Maluf2013} uses this scalar $T$ as the gravitation Lagrangian density.
And the minimal matter-gravity coupling $f(T)$ gravity
extends it to an arbitrary function of $T$.
The nonminimal coupling $f(T)$ gravity\cite{Harko2014a}
further extends the matter Lagrangian term $\mathcal{L}_\text{m}$
into $(1+f_2(T))\mathcal{L}_\text{m}$.
To be unifiable, we write the action in the following form:
\begin{equation}
	S=-\frac1{16\pi G}\int|e|\left( 1+f_1(T) \right)T\diff^4x+\int|e|\left( 1+f_2(T) \right)\mathcal{L}_\text{m}\diff^4x.
	\label{actionnonmin}
\end{equation}
And the actions of TEGR and minimal coupling $f(T)$ gravity correspond to the
$f_1,f_2=0$ and $f_2=0$ cases of Eq.\eqref{actionnonmin}, respectively.
With the action principle applied on Eq.\eqref{actionnonmin}
with respect to the tetrad field,
the field equation is then given by
\begin{equation}
		\frac4{|e|}f\partial_\beta(|e|S_\sigma^{\:\:\alpha\beta}e_a^{\:\sigma})+4e_a^{\:\sigma}S_\sigma^{\:\:\alpha\beta}\partial_\beta f+4fS_\rho^{\:\:\alpha\sigma}T^\rho_{\:\:\sigma\beta}e_a^{\:\beta}+(1+f_1)Te_a^{\:\alpha}=-16\pi G(1+f_2)\mathcal{T}_\beta^{\:\alpha}e_a^{\:\beta},
	\label{eom}
\end{equation}
where $f\equiv1+f_1(T)+f_1'(T)T-16\pi Gf_2'(T)\mathcal{L}_\text{m}$, and $\mathcal{T}_\beta^{\:\alpha}$
is the energy-momentum tensor of matter given by
\begin{equation}
	\frac{\delta(|e|\mathcal{L}_\text{m})}{\delta e^a_{\:\alpha}}=-|e|\mathcal{T}_\beta^{\:\alpha}e_a^{\:\beta}.
	\label{emtensor}
\end{equation}
And it takes the usual form for perfect fluid
\begin{equation}
	\mathcal{T}_{\mu\nu}=pg_{\mu\nu}-(\rho+p)u_\mu u_\nu,
	\label{emtensor1}
\end{equation}
where $p$ and $\rho$ are the pressure and energy density of the matter, respectively,
and $u^\mu$ is the 4-velocity.

The covariant derivative (related to the Levi-Civit{\'a} connection) of Eq.\eqref{eom} gives
\begin{equation}
	\nabla^\nu\mathcal{T}_{\mu\nu}=-\frac{f_2'(T)\nabla^\nu T}{1+f_2(T)}(\mathcal{T}_{\mu\nu}+g_{\mu\nu}\mathcal{L}_\text{m}).
	\label{covder}
\end{equation}
This suggests that the energy-momentum tensor is no longer conservative.
However,
contracting Eq.\eqref{covder} with $u^\mu$, we have
\begin{equation}
	\begin{split}
		u^\mu\nabla^\nu\mathcal{T}_{\mu\nu}=&\frac{f_2'(T)\nabla^\nu T}{1+f_2(T)}(\rho-\mathcal{L}_\text{m})\\
		=&-u^\mu\nabla_\mu\rho-(\rho+p)\nabla_\mu u^\mu.
	\end{split}
	\label{mattere}
\end{equation}
If we take the matter Lagrangian density to be
$\mathcal{L}_\text{m}=\rho$\cite{Gron2007,Bertolami2008,Minazzoli2012,Harko2014},
then in cosmological cases,
as discussed in Ref.\cite{Feng2015},
Eq.\eqref{mattere} will return to the usual form of conservation law of matter in
Friedmann-Lema{\^i}tre-Robertson-Walker (FLRW) metric.

\subsection{Observationally tested models}
Consider a flat FLRW universe with metric in Cartesian coordinates
\begin{equation}
	g_{\mu\nu}=\text{diag}(1,-a^2(t),-a^2(t),-a^2(t)),
	\label{FRWmetric}
\end{equation}
with scale factor $a(t)$.
It is found\cite{Tamanini2012} that for this form of metric,
the diagonal vierbein $e^a_{\:\mu}=\text{diag}(1,a(t),a(t),a(t))$
is a viable choice for $f(T)$ gravities.
And then the torsion scalar $T=-6H^2$,
where $H=\dot{a}/a$ is the Hubble parameter,
and the overdot denotes derivative with respect to time.
The 00 entry of Eq.\eqref{eom} then reads
\begin{equation}
	12fH^2-6\left( 1+f_2 \right)H^2=16\pi G\left( 1+f_2 \right)\rho.
	\label{coseom}
\end{equation}
When $f_1,f_2=0$, i.e. $f=1$,
Eq.\eqref{coseom} gives the usual Friedmann's equation in GR or TEGR,
and when $f_2=0$, Eq.\eqref{coseom} gives the field equation of minimal coupling $f(T)$ gravities.

For another equation to determine the system,
one can either use the rest components of Eq.\eqref{eom} or
the evolution of matter.
Here we take the latter scheme.
From Eq.\eqref{mattere},
if we take $\mathcal{L}_\text{m}=\rho$
and assume the matter is dust-like $p\ll\rho$,
we can have the familiar evolution of matter $\rho=\rho_0 (a_0/a)^{3}$,
where $\rho_0$ and $a_0$ are the energy density of matter and
scale factor at present time, respectively.
Usually $a_0$ is set to be 1.

In Ref.\cite{Feng2015},
we have constructed two concrete models of
nonminimal coupling $f(T)$ gravities
for cosmological fitting.
For completeness,
here we consider one additional model
of minimal coupling $f(T)$ gravity.
They are listed as follows:
\begin{itemize}
	\item Model I: $f_1=\frac{12BH_0^4}{T^2}$, $f_2=0$;
	\item Model II: $f_1=H_0(-T)^{-\frac12}$, $f_2=\frac{-2AH_0^2}{\Omega_\text{m0}T}$;
	\item Model III: $f_1=\frac{12BH_0^4}{T^2}$,  $f_2=\frac{-2AH_0^2}{\Omega_\text{m0}T}$,
\end{itemize}
where $A,B$ are the parameters of the models,
$H_0$ is the Hubble parameter at present time and
$\Omega_\text{m0}=\frac{8\pi G\rho_0}{3H_0^2}$ is the matter density parameter.
Model I represents the case of minimal coupling $f(T)$ gravity,
while Model II and III are the models of nonminimal coupling $f(T)$ gravity
constructed and fitted in Ref.\cite{Feng2015}.
Then Eq.\eqref{coseom} can be unifiably written as
\begin{equation}
	\left( \frac{\dot a}a \right)^2=\Omega_\text{m0}H_0^2a^{-3}+\frac{\left( B+Aa^{-3} \right)H_0^4}{\left( \frac{\dot a}a \right)^2}.
	\label{flatunieom}
\end{equation}
Denoting $E=H/H_0$, we have
\begin{equation}
	E^2=\Omega_\text{m0}a^{-3}+Aa^{-3}E^{-2}+BE^{-2}
	\label{unieom}
\end{equation}
with $1=\Omega_\text{m0}+A+B$.
Model I, II, III correspond to the cases
$A=0$, $B=0$ and $A,B\ne0$, respectively.

We use the data set of SNeIa, BAO and CMB,
the same as Ref.\cite{Feng2015},
to find the best-fit values of parameters for Model I,
and retain the fitting results of Model II and III,
listed in Table \ref{bestfit}.

\begin{table}[htbp]
  \centering
  \caption{Best-fit parameters for the models}
    \begin{tabular}{cccc}
    \hline
    \hline
          & \multicolumn{3}{c}{models} \\
\cline{2-4}    parameters & Model I & Model II & Model III \\
    \hline
    $\Omega_\text{m0}$   & $0.274\pm0.008$ & $0.367\pm0.011$ & $0.302\pm0.011$ \\
    $H_0$    & $73.73\pm0.79$ & $60.97\pm0.44$ & $68.54\pm1.27$ \\
    $A$   &   -    &   $0.633\pm0.011$    & $0.188\pm0.048$ \\
    $\chi_\text{min}^2/\text{d.o.f.}$     & 707.455/739 & 729.429/739 & 683.846/739 \\
    \hline
    \hline
    \end{tabular}%
  \label{bestfit}%
\end{table}%

\section{Spherical top-hat collapse}
\label{collapse}
The classical spherical top-hat collapsing formalism
considers a spherical region with uniformly perturbed energy density
immersed in the homogeneous universe.
The initial magnitude of the perturbation is
denoted as $\delta_\text{i}$,
and it occurs at the redshift $z_\text{i}$,
where the subscript "i" hereinafter indicates
the initial evaluation of the quantities at this very moment.
The initial radius of the overdense region
and the density of the universe are denoted as
$R_\text{i}$ and $\rho_\text{i}$, respectively.
Thus the initial density of the overdense region is $\rho_\text{i}(1+\delta_\text{i})$
and the scale factor at that time is $a_\text{i}=1/(1+z_\text{i})$.

This region then detaches from the rest of the universe
and evolves on its own like a non-spatial-flat universe
\cite{Gunn1972,Padmanabhan1993,Mota2004,Abramo2009}.
For this reason,
we need the equation of a spatial curved universe
besides the flat background evolution.
For a universe with spatial curvature $k$,
the torsional formalism gives the torsion scalar\cite{Tamanini2012}
\begin{equation}
	T=-6H^2+\frac{6k}{a^2}.
	\label{kT}
\end{equation}
And the 00 entry of Eq.\eqref{eom} is
\begin{equation}
	\left( 2f-1-f_1 \right)\left( 6H^2-\frac{6k}{a^2} \right)=16\pi G\left( 1+f_2 \right)\rho.
	\label{curvcoseom}
\end{equation}
When $k=0$, Eq.\eqref{coseom} is recovered.
For the three models listed above,
Eq.\eqref{curvcoseom} is then
\begin{equation}
	\left( \frac{\dot a}a \right)^2-\frac{k}{a^2}=\Omega_\text{m0}H_0^2a^{-3}+\frac{\left( B+Aa^{-3} \right)H_0^4}{\left( \frac{\dot a}a \right)^2-\frac{k}{a^2}}.
	\label{unieom0}
\end{equation}
Denoting $H_\text{i}=H(a_\text{i})$,
$\tilde k=\frac{k}{a_\text{i}^2H_\text{i}^2}$,
$\Omega_\text{mi}=\frac{8\pi G\rho_0}{3H_\text{i}^2a_\text{i}^3}$,
$\tilde A=A\frac{H_0^4}{a_\text{i}^{3}H_\text{i}^{4}}$ and
$\tilde B=B\frac{H_0^4}{H_\text{i}^{4}}$, we have
\begin{equation}
	\left(\frac{\dot a}{H_\text{i}a_\text{i}}\right)^2-\tilde k=\Omega_\text{mi}\frac{a_\text{i}}{a}+\frac{\tilde A\left( \frac{a}{a_\text{i}} \right)+\tilde B\left( \frac{a}{a_\text{i}} \right)^4}{\left(\frac{\dot a}{H_\text{i}a_\text{i}}\right)^2-\tilde k}.
	\label{unieom1}
\end{equation}
This is the equation governing the scale factor $a$
in a spatial curved universe.
Therefore the equation for the radius $R$ of the overdense region is in the similar form,
\begin{equation}
	\left(\frac{\dot R}{h_\text{i}R_\text{i}}\right)^2-\tilde k=\Omega_\text{mi}\left( 1+\delta_\text{i} \right)\frac{R_\text{i}}{R}+\frac{\tilde A\left( \frac{R}{R_\text{i}} \right)+\tilde B\left( \frac{R}{R_\text{i}} \right)^4}{\left(\frac{\dot R}{h_\text{i}R_\text{i}}\right)^2-\tilde k},
	\label{reom}
\end{equation}
with
\begin{equation}
	1-\tilde k=\Omega_\text{mi}\left( 1+\delta_\text{i} \right)+\frac{\tilde A+\tilde B}{1-\tilde k},
	\label{rnorm}
\end{equation}
where $h=\dot R/R$ is the expansion rate of the region
and initially $h_\text{i}=H_\text{i}$.

\begin{figure}[!htp]
	\centering
	\includegraphics[width=0.9\textwidth]{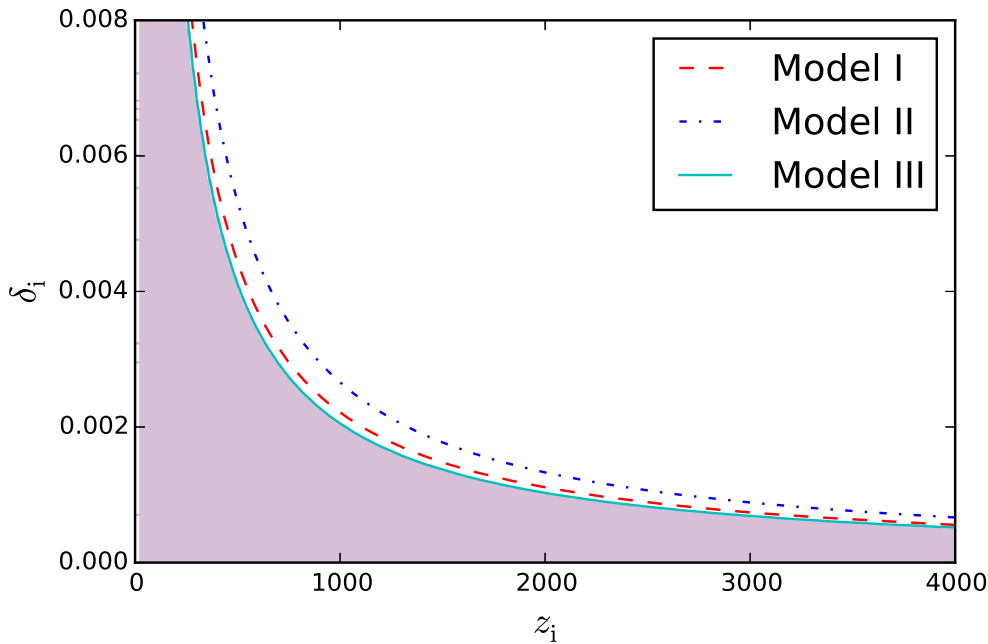}
	\caption{The constraints of $z_\text{i}$ and $\delta_\text{i}$
for Model I(dashed), II(dashdot) and III(solid),
where the models have taken the best-fit parameters
listed in Table \ref{bestfit}.
The shaded area is forbidden for
not being able to turn the region
from expanding to collapsing.}
	\label{tacon}
\end{figure}

As the evolution continues,
the region may stop expanding and start collapsing due to the overdensity.
At the turn-around point,
if it ever occurs,
the radius of the region reaches its maximum $R_\text{max}$
and $\dot R=0$.
Denoting $\beta\equiv R_\text{max}/R_\text{i}$,
we can rewrite Eq.\eqref{reom} at the turn-around point as
\begin{equation}
	\tilde B\beta^5+\tilde A\beta^2-\tilde k^2\beta-\tilde k\Omega_\text{mi}(1+\delta_\text{i})=0.
	\label{quintic}
\end{equation}
It is a quintic equation of $\beta$.
The region will turn from expanding to collapsing
if only Eq.\eqref{quintic} has real positive root(s).
According to the Passare-Tsikh solution\cite{Passare2004}
of the principal quintic
\begin{equation}
	m x^5+n x^2+x+1=0,
	\label{pquintic}
\end{equation}
Eq.\eqref{quintic} has real positive root(s) if and only if (See Appendix \ref{appendix})
\begin{equation}
	3125|m|^2-256|m|+108|n|^5-27|n|^4+625|m||n|-2250|n|^2|m|<0
	\label{condition}
\end{equation}
where
\begin{equation}
	m=-\frac{\tilde B\Omega_\text{mi}^4(1+\delta_\text{i})^4}{\tilde k^6},\quad n=-\frac{\tilde A\Omega_\text{mi}(1+\delta_\text{i})}{\tilde k^3}.
	\label{mn}
\end{equation}
For Model I ($A=0$),
the quadratic term of Eq.\eqref{pquintic} vanishes,
while the discriminant \eqref{condition} is still valid
and reduces to
$3125|m|-256<0$, namely
\begin{equation}
	\frac{\left( 1+z_\text{i} \right)^{12}\left( 1+\delta_\text{i} \right)^{3}}{H_\text{i}^{12}\tilde k^6}<\frac{256}{3125 BG^4H_0^{12}\Omega_\text{m0}^4}.
	\label{disI}
\end{equation}
For Model II ($B=0$),
the quintic term of Eq.\eqref{pquintic} vanishes
and the equation reduces to a quadratic one.
And Eq.\eqref{condition} gives the same inequality
as the discriminant of the quadratic:
$4|n|-1<0$, namely
\begin{equation}
	-\frac{\left( 1+z_\text i \right)^3\left( 1+\delta_\text i \right)}{H_\text i^2\tilde k^3}<\frac1{4A\Omega_\text{m0}H_0^2}.
	\label{disII}
\end{equation}
For Model III ($A,B\ne0$),
similar inequality can be obtained,
which is omitted here for its lengthiness.
In Eqs.\eqref{disI}, \eqref{disII} and the corresponding inequality for Model III,
$H_\text{i}$ and $\tilde k$ depend on $z_\text{i}$ and $\delta_\text{i}$
according to Eqs.\eqref{unieom} and \eqref{rnorm}, respectively.
And thus for given $A$ and $B$,
these inequalities give constraints on the time ($z_\text{i}$)
and magnitude ($\delta_\text{i}$)
of the initial perturbation.
For the three models listed above,
we have plotted these constraints in Fig.\ref{tacon},
where the models have taken the best-fit parameters listed in Table \ref{bestfit}.
Any pairs of $\left( z_\text{i},\delta_\text{i} \right)$ above
the illustrated lines can guarantee real positive root(s)
of Eq.\eqref{quintic},
namely make sure that the overdense region will
eventually collapse and form structure.
The result is not surprising in that
a perturbation that either is too weak (low $\delta_\text{i}$)
or occurs too late (low $z_\text{i}$,
and such that the matter of the universe is too thin)
will not be able to provide a self-gravity strong enough
to slow down the expanding and turn the region into collapsing.
Compared to the minimal coupling Model I,
the constraint for Model II is more stringent,
while the one for Model III is slightly looser.

Note that
if $\tilde A=\tilde B=0$ in Eq.\eqref{quintic},
$\beta=-\Omega_\text{i}(1+\delta_\text{i})/\tilde k>0$
is automatically the physical root we need.
That is,
for Einstein-de Sitter universe in GR or TEGR,
as shown in many standard cosmology textbooks,
these constraints for $\delta_\text{i}$ and $z_\text{i}$ do not exists
and collapse will always happen eventually due to
the lack of persistent driving source of the expansion.

\begin{figure}
	\centering
	\includegraphics[width=0.9\linewidth]{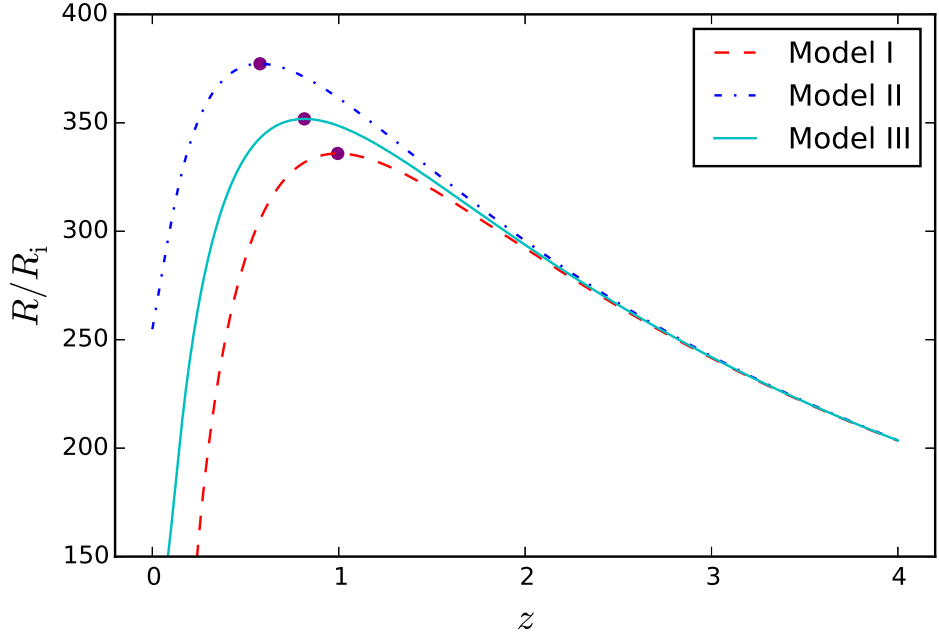}
	\caption{The evolution of $R/R_\text{i}$ of all three models,
where $z_\text{i}=1200$, $\delta_\text{i}=0.003$,
and the models have taken
the best-fit parameters from Table \ref{bestfit}.}
	\label{revol}
\end{figure}

\begin{figure}
	\centering
	\includegraphics[width=0.9\linewidth]{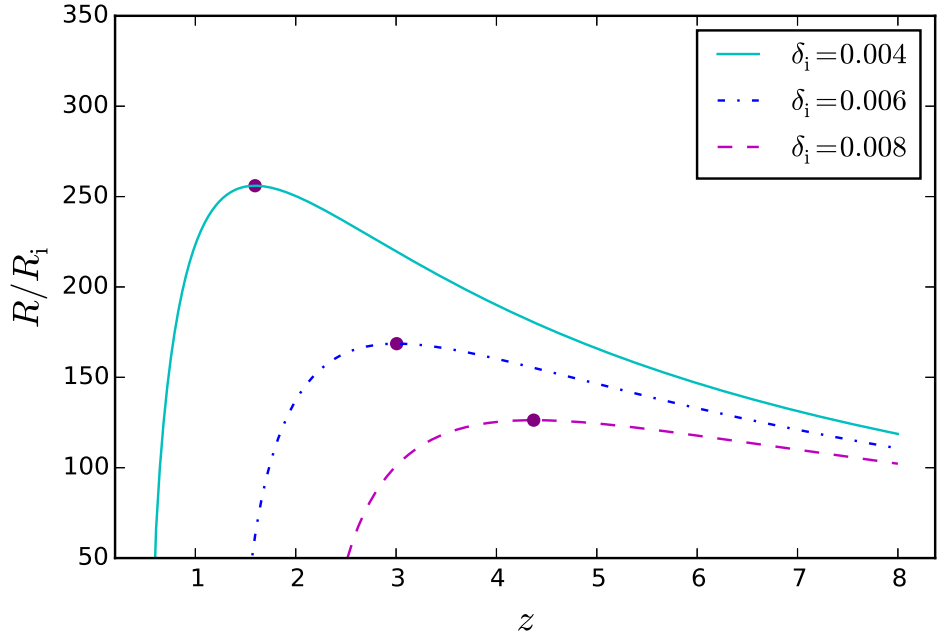}
	\caption{The evolution of $R/R_\text{i}$ for different $\delta_\text{i}$ in model III,
where $z_\text{i}=1200$,
and the models have taken
the best-fit parameters from Table \ref{bestfit}.}
	\label{m3di}
\end{figure}

For a given model and initial perturbation,
Eq.\eqref{reom} determines the evolution of the radius of the overdense region,
which can be carried out numerically.
The evolution of $R/R_\text{i}$ of all three models are plotted in Fig.\ref{revol},
where we have set $z_\text{i}=1200$, $\delta_\text{i}=0.003$,
and taken the best-fit parameters from Table \ref{bestfit}. The variation of $\delta_\text{i}$ will also alter the evolution of $R/R_\text{i}$, which is illustrated in Fig.\ref{m3di} for Model III.

\begin{figure}[!htp]
	\centering
	\includegraphics[width=0.9\linewidth]{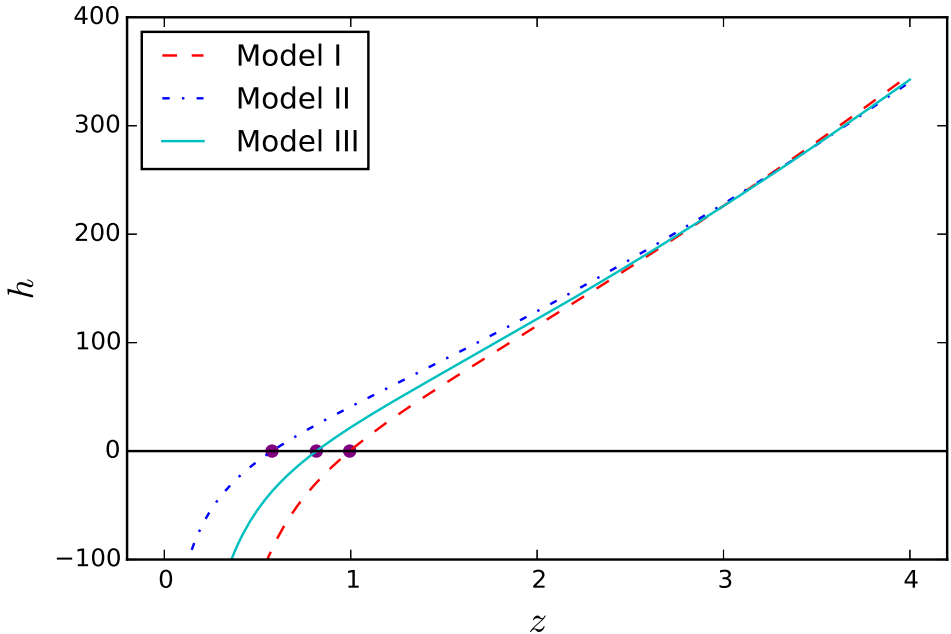}
	\caption{The evolution of $h=\dot R/R$ of all three models,
where $z_\text{i}=1200$, $\delta_\text{i}=0.003$,
and the models have taken
the best-fit parameters from Table \ref{bestfit}.}
	\label{hevol}
\end{figure}

We have also calculated the evolutions of the expansion rate $h=\dot R/R$ of the region,
and plotted them in Fig.\ref{hevol}.
The markers in Figs.\ref{revol}-\ref{hevol}
correspond to the turn-around points of the models,
the evaluations of which are listed in Table \ref{betaalpha}.
One can see that
compared to the minimal coupling model (Model I),
the collapsing processes are slower
in the nonminimal coupling models (Model II and III),
thus the turn-arounds
happen at lower redshift $z_\text{ta}$
and consequently larger $\beta=R_\text{max}/R_\text{i}$ are reached.
In the case of Model II where effective dark energy
is entirely attributed to the nonminimal torsion-matter coupling,
the collapsing is the slowest among the three,
and largest $\beta$ is reached at latest time $z_\text{ta}$.

\begin{figure}[!htp]
	\centering
	\includegraphics[width=0.9\linewidth]{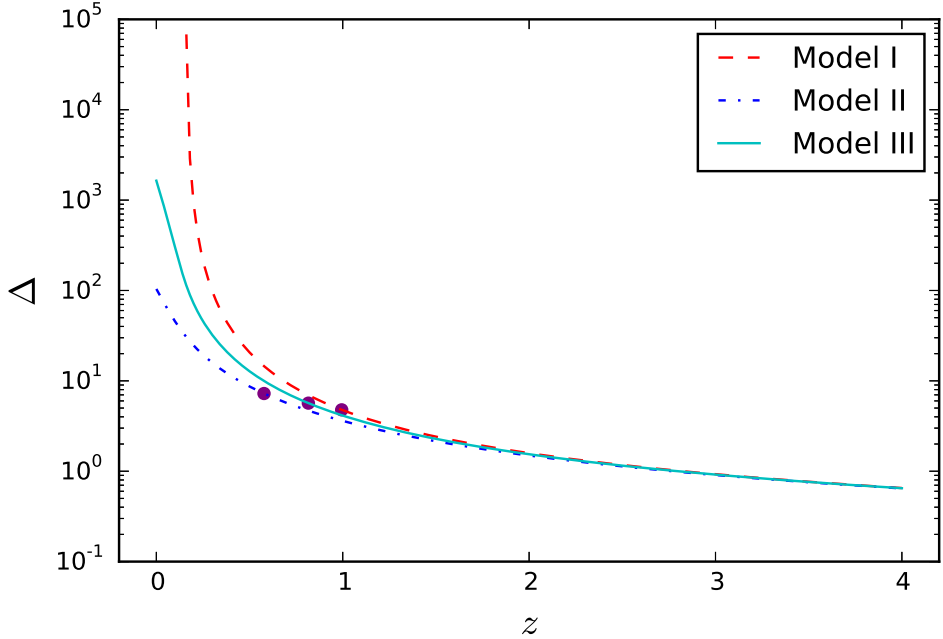}
	\caption{The evolution of $\Delta=\rho_\text{c}/\rho_\text{u}$ of all three models,
where $z_\text{i}=1200$, $\delta_\text{i}=0.003$,
and the models have taken
the best-fit parameters from Table \ref{bestfit}.}
	\label{devol}
\end{figure}

Moreover, it is also worth considering
the evolution of the density contrast between
the collapsing region and the background universe.
Denoting the density of the universe as
$\rho_\text{u}=\rho_\text{i}(a_\text{i}/a)^3$ and
that of the collapsing region as
$\rho_\text{c}=\rho_\text{i}(1+\delta_\text{i})(R_\text{i}/R)^3$,
one then can define the density contrast $\Delta$ by
$\rho_\text{c}=\rho_\text{u}\Delta$.
Utilizing the evolutions of $R$ from Eq.\eqref{reom} and
$a$ from Eq.\eqref{flatunieom},
we have carried out the evolution of $\Delta$,
and illustrated in Fig.\ref{devol}.

\section{Virialization of the collapse}
\label{virial}
In Fig.\ref{devol},
one can see that after the turn-around point,
the region governed by Eq.\eqref{reom}
will soon collapse into a singularity,
which, however,
is not the actual fate for all the collapsing structures.
As the clustering continues,
the kinetic energy of the matter
will be no longer negligible,
which will eventually take the system to an equilibrium
and stop the collapsing.
This highly non-linear process is simplified
and described by the virialization of a self-gravity system.

The relativistic virial theorem starts from the
Boltzmann's equation for collisionless particles
(see, e.g., \cite{Jackson1970,Debbasch2009}):
\begin{equation}
	u^\mu\partial_\mu F_\text{B}+\frac{\diff p^\mu}{\diff\tau}\frac{\partial F_\text{B}}{\partial p^\mu}=0,
	\label{boltzmann}
\end{equation}
where $u^\mu$ is the 4-velocity, $p^\mu$ is the 4-momentum,
$\diff p^\mu/\diff\tau$ is the 4-force,
and $F_\text{B}$ is the Boltzmann distribution function
on the 6-dimensional phase space $\{\vec{x},\vec{p}\}$.
Assume that $F_\text{B}$ vanishes sufficiently rapidly
as the velocities tend to infinity,
Eq.\eqref{boltzmann} will lead to the virial theorem
(see, e.g., \cite{Gunn1972,Bohmer2008,Javadinezhad2016}):
\begin{equation}
	2\mathcal{E}_\text{K}+\int\vec{f}\cdot\vec{x}\diff V=0,
	\label{virialth}
\end{equation}
where $\mathcal{E}_\text{K}$ is the kinetic energy of the system
and $\vec{f}$ is the force that the system imposes on
the local volume element $\diff V$ located at $\vec{x}$.
The integral is taken over the whole system.

Usually, the force $\vec f$ or the 4-force $\diff p^\mu/\diff\tau$
should be obtained from the geodesic equation,
or the non-geodesic equation for the nonminimal coupling cases
(see, e.g. \cite{Harko2014,Harko2014a,Feng2015}),
but that requires full knowledge of the spherical symmetric solution
to the field equation of the modified gravity,
which we do not currently have.
On the other hand,
the Newtonian-like gravity force and potential
are connected to the cosmology.
In GR, this argument is originated by Milne\cite{MILNE1934},
and presented in many standard cosmology textbooks.

Consider a comoving test particle located at $r=r_\text{i}a/a_\text{i}=r_0a$,
where $r_0=r_\text{i}/a_\text{i}$ is its comoving position,
then Eq.\eqref{flatunieom} can be written in terms of $r$ as
\begin{equation}
	\begin{split}
		\left( \frac{\dot r}r \right)^2=&\frac{8\pi G}3\frac{\rho_0r_\text{i}^3}{a_\text{i}^3r^3}+\frac{AH_0^4r_\text{i}^3}{a_\text{i}^3r^3}\left( \frac r{\dot r} \right)^2+BH_0^4\left( \frac r{\dot r} \right)^2\\
		=&\frac{8\pi G\rho_\text{i}}3\left( \frac{r_\text{i}}{r} \right)^3+\left[\tilde A\left( \frac{r_\text{i}}r \right)^3+\tilde B\right]h_\text{i}^4\left( \frac r{\dot r} \right)^2,
	\end{split}
	\label{comove}
\end{equation}
which is the equation of motion of the test particle.
Denoting the mass inside a spherical region of radius $r$ as $m=4\pi r_\text{i}^3\rho_\text{i}/3$,
we have
\begin{equation}
	\frac{\left( \dot r \right)^2}2=\frac{Gm}{2r}+\frac12\sqrt{\left( \frac{Gm}r \right)^2+h_\text{i}^4\left( \tilde Ar_\text{i}^3r+\tilde Br^4 \right)}.
	\label{comove1}
\end{equation}
In classic analogy, $\dot r$ is the velocity of the particle,
and the left hand side of Eq.\eqref{comove1}, $(\dot r)/2$,
is the kinetic energy density of it.
We then can identify the right hand side of Eq.\eqref{comove1} as the gravitational potential per unit mass at $r$
\begin{equation}
	V(r)=-\frac{Gm}{2r}-\frac12\sqrt{\left( \frac{Gm}r \right)^2+h_\text{i}^4\left( \tilde Ar_\text{i}^3r+\tilde Br^4 \right)},
	\label{potentialr}
\end{equation}
where $m=4\pi r_\text{i}^3\rho_\text{i}/3=4\pi r^3\rho/3$.
When $\tilde A=\tilde B=0$, it is obviously the Newtonian potential.

And the force per unit mass acting on the volume element at $r$ is
\begin{equation}
	\begin{split}
		f(r)=&-\frac{\diff V(r)}{\diff r}\\
		=&-\frac{Gm}{2r^2}+\frac{-2\frac{G^2m^2}{r^3}+h_\text{i}^4\left( 4\tilde Br^3+\tilde Ar_\text{i}^3 \right)}{4\sqrt{\frac{G^2m^2}{r^2}+h_\text{i}^4\left( \tilde Br^4+\tilde Ar_\text{i}^3r \right)}}.
	\end{split}
	\label{forcer}
\end{equation}
Hence the virial theorem is
\begin{equation}
	\begin{split}
		2\mathcal{E}_\text{K}=&-\int_0^Rf(r)r4\pi r^2\rho\diff r\\
		=&-\frac{9M}{20R}\frac{h_\text{i}^4R^3\left( 2\tilde BR^3+\tilde AR_\text{i}^3 \right)}{\sqrt{G^2M^2+h_\text{i}^4R^3\left( \tilde BR^3+\tilde AR_\text{i}^3 \right)}}\\
		&+\frac{3GM^2}{10R}+\frac{3M}{10R}\sqrt{G^2M^2+h_\text{i}^4R^3\left( \tilde AR_\text{i}^3+\tilde BR^3 \right)}.
	\end{split}
	\label{virialend}
\end{equation}

From Eq.\eqref{potentialr},
if a spherical system has a radius $R$ and mass $M=4\pi R^3\rho/3$,
the total potential energy of it is
\begin{equation}
	\begin{split}
		V_\text{tot}(R)=&\int_0^RV(r)4\pi r^2\rho\diff r\\
		=&-\frac{3GM^2}{10R}-\frac{3M}{10R}\sqrt{G^2M^2+h_\text{i}^4R^3\left( \tilde AR_\text{i}^3+\tilde BR^3 \right)},
	\end{split}
	\label{potentialtot}
\end{equation}
where $R_\text{i}^3=R^3\rho/\rho_\text{i}$.

Although usually the energy conservation is considered violated
in the nonminimal coupling cases due to
the extra force and non-geodesic movement
(see Eq.\eqref{covder} or, e.g., \cite{Harko2014,Harko2014a,Feng2015}),
here the energy conservation is still hold in form during the collapse,
since we have put the effect of the extra force into
the gravitation potential.
And hence, we have
\begin{equation}
	\mathcal{E}_\text{K,vir}+V_\text{tot}(R_\text{vir})=V_\text{tot}(R_\text{max}),
	\label{conservation}
\end{equation}
where $R_\text{vir}$ is the radius of the collapsing region
when it reaches virialization.

Denote $\alpha=R_\text{vir}/R_\text{i}$,
Eq.\eqref{conservation} gives
\begin{equation}
	\begin{split}
		&\frac{\Omega_\text{mi}(1+\delta_\text{i})}{2\beta}+\frac1{\beta}\sqrt{\frac{\Omega_\text{mi}^2(1+\delta_\text{i})^2}{4}+\beta^3(\tilde A+\tilde B\beta^3)}\\
		=&\frac{\Omega_\text{mi}(1+\delta_\text{i})}{4\alpha}+\frac1{2\alpha}\sqrt{\frac{\Omega_\text{mi}^2(1+\delta_\text{i})^2}{4}+\alpha^3(\tilde A+\tilde B\alpha^3)}\\
		&+\frac{3\alpha^2(2\tilde B\alpha^2+\tilde A)}{4\sqrt{\frac{\Omega_\text{mi}^2(1+\delta_\text{i})^2}4+\alpha^3(\tilde A+\tilde B\alpha^3)}},
	\end{split}
	\label{alpha}
\end{equation}
where $\beta=R_\text{max}/R_\text{i}$.
Using the numerical evaluation of $\beta$ from Eq.\eqref{quintic}
and $\alpha$ from Eq.\eqref{alpha},
one can obtain the collapse factor $\lambda=R_\text{vir}/R_\text{max}=\alpha/\beta$.
Furthermore, one can in turn calculate the redshifts and density contrasts
at turn-around $(z_\text{ta},\:\Delta_\text{ta})$
and virialization $(z_\text{vir},\:\Delta_\text{vir})$
from the evolutions plotted in Figs.\ref{revol} and \ref{devol}.
The evaluations at the turn-around and virialization for the models with
$z_\text{i}=1200$, $\delta_\text{i}=0.003$
are listed in Table \ref{betaalpha}.
The corresponding results for Einstein de Sitter universe
are also listed for comparison.
One can see that the results for the minimal coupling model (Model I)
are relatively closer to the E-dS model,
while the nonminimal coupling models (Model II and III)
virialize slower.
The virialization for Model II even happens in the future
$(z_\text{vir}<0)$ for the given setting.

\begin{table}[htbp]
  \centering
  \caption{Turn-around and virialization of the models}
    \begin{tabular}{ccccc}
    \hline
    \hline
          & \multicolumn{4}{c}{models} \\
\cline{2-5}    & Model I & Model II & Model III & E-dS\\
    \hline
    $z_\text{ta}$   & $0.994$ & $0.578$ & $0.816$ & $1.031$\\
    $R_\text{max}/R_\text{i}$    & $335.864$ & $377.229$ & $351.814$ & $334.333$\\
    $\Delta_\text{ta}$   &  $ 5.781$    &  $ 8.242 $   & $6.666$ & $5.551$\\
    \hline
    $z_\text{vir}$   & $0.258$ & $-0.189$ & $0.110$ & $0.363$\\
    $R_\text{vir}/R_\text{i}$    & $167.245$ & $176.837$ & $171.118$ & $167.167$\\
    $\Delta_\text{vir}$   &  $186.470$    &  $ 589.284 $   & $253.331$ & $146.841$\\
    \hline
    $\lambda$& $0.498$ & $0.469$ & $0.486$ & $0.5$\\
    \hline
    \hline
    \end{tabular}%
  \label{betaalpha}%
\end{table}%

\section{Conclusion and discussion}
\label{conclusion}
Most structures in our universe
such as stars, galaxies and clusters of galaxies
are formed from the nonlinear evolution of perturbations during the post-recombination epoch. Ideally, we should hope to investigate the problem by combining $f(T)$ models of large-scale structure formation with hydrodynamic codes which follow the dynamical and thermal histories of the diffuse intergalactic gas. A fully relativistic treat of this non-linear perturbation is not currently available. The intrinsically non-linear nature of the processes is usually handled by N-body simulation, which can be cumbersome and time-consuming, and is not practical for one to use to study different gravitation models. However, the spherical top-hat profile is the simplest analytic model which can be used to study the non-linear growth of spherically symmetric density perturbation and structure formation in $f(T)$ gravities. Though density perturbation will not be spherically symmetric in most of the realistic cases, the spherical top-hat profile can give significant enlightenment for the true picture of physics. Anyhow, the top-hat profile is a useful means for the non-linear evolution of perturbation. In this paper, we have studied the non-linear growth of perturbation in $f(T)$ gravities, including both minimal and nonminimal coupling cases. And then we looked into the virialization of the collapsing region after the turn-around point. Our main results are as follows:

\begin{enumerate}[(i)]
	\item The numerical calculation of redshift $z_{\text{ta}}$ at the turn-around point indicates incontestably that galaxies and other large-structure features must have formed relatively late in the history of the universe in three $f(T)$ models. If we take appropriate initial conditions $\delta_\text{i}$ at $z=z_\text{i}$, the value of $z_{\text{ta}}$ will not conflict with the late time formation of the structures(see Fig.\ref{m3di}). As a matter of fact, the population of galaxies and quasars have evolved dramatically over the redshift range $0<z<6$.

\item The perturbations were still in the linear regime at $z\sim 1000$, but as they entered upon the later phase of the post-recombination era, their evolution became non-linear. In Ref.\cite{Feng2015}, we have given the relation between $a(t)$ and $t$ at the post-recombination era for the $f(T)$ cosmological models (see Eqs. (45)-(47) in Ref.\cite{Feng2015}). Thereby, we can use the linear evolution equation of density perturbation and get the result of $\delta(z_{\text{ta}})$ which is far more smaller than that of top-hat profile. By the time the perturbed sphere had stopped expanding, its density in the three models is already 5.781, 8.242 and 6.666 times greater than that of the background density. These pictures are consistent with the primitive spherical top-hat collapse (E-dS model).

\item The existence of turn-around point is the crux of matter in the non-linear collapse of density perturbations. The perturbation can always reaches maximum radius for the primitive top-hat model. There are constraints of initial value $\delta_\text{i}$ at $z_\text{i}$ in three $f(T)$ models (see Fig.\ref{tacon}), which reinforce our intuition that in an accelerating expanding universe, a magnitude of perturbation  $\delta_\text{i}$ that is too weak will not be able to turn the overdense region from expanding to collapsing. In Ref. \cite{Feng2015}, we have given $z_{\text{crit}}=0.3298$ and 0.6694 for Model II and Model III, respectively. Here, $z_{\text{crit}}$ is defined as the critical values of the redshift through which the universe changes to the acceleration phase from the deceleration one. Therefore, we have a reasonable explanation for our numerical results. We find that compared to Model III, the collapsing processes are slower since $z_{\text{crit}}$ is smaller in Model II (see Fig.\ref{revol}). Furthermore, the turn-around happens at lower redshift $z_{\text{ta}}$ and consequently larger $R_{\text{max}}/R_\text{i}$ is reached.

\item Outwardly, the subsequent collapse occurs very rapidly (see Fig. \ref{revol}) after the redshift of $z_{\text{ta}}$ and the spherical perturbed region collapses ultimately to a black hole. In reality, however, it is much more likely to form some sort of bound object. If the density becomes high enough, we cannot neglect the internal pressure in the overdense region. As the gas cloud collapses, its temperature increases until pressure gradients becomes sufficient to balance the attractive force of gravitation. The ultimate result is a system which satisfies the virial theorem. By the numerical calculation, we have obtained the density contrast between the collapsing region and the background universe by the time the region is virialized (see Table 2). Since the critical value $z_{\text{crit}}$ of Model II is smaller than that of Model III, we have a good reason to expound the slower virialized process for Model II.

\item  The numerical results show that nonminimal coupling models collapse slower than the minimal coupling model, which may be an indication for alleviating the small scale problems
    \cite{Moore1994,Boylan-Kolchin2012}.

\item The virialization for Model II happens in the future ($z_{\text{vir}}=-0.189$) for the given setting $z_\text{i}=1200$ and $\delta_\text{i}=0.003$. This is not strange because the spherical top-hat collapse is a highly idealised picture. The better approximation is to assume that the collapsing regions are ellipsoidal with three unequal principal axes. In this model, the ellipsoid collapses to a 'pancake' which is the collapsing result of the non-linear regime in the more general case.
  \end{enumerate}

To sum up, the spherical top-hat collapse is a qualitative analysis which successfully expounds the aspects of non-linear evolution in $f(T)$ gravities. But this is a highly idealised description, so we do not really expect that it is quantitatively consistent with the process of formation of bound structures in the expanding universe. For example, the observation of CMB requires $\Delta T/T\sim 10^{-5}$, which is corresponding to $\delta\rho/\rho\sim 10^{-5}$ at $z=1200$. In other words, the spherical top-hat collapse is only a toy model. For comparison with the observation, we should study further the pancake model and N-body simulations in $f(T)$ gravities. We will look into these issues in the future works.
\bibliography{ref}

\providecommand{\href}[2]{#2}\begingroup\raggedright\begin{thebibliography}{10}

\bibitem{Capozziello2011}
S.~Capozziello and M.~De~Laurentis, \emph{Extended theories of gravity},
  \href{http://dx.doi.org/10.1016/j.physrep.2011.09.003}{\emph{Phys. Rep.} {\bf
  509} (2011) 167--321}.

\bibitem{Sotiriou2010}
T.~P. Sotiriou and V.~Faraoni, \emph{$f(\mathit{R})$ theories of gravity},
  \href{http://dx.doi.org/10.1103/revmodphys.82.451}{\emph{Rev. Mod. Phys.}
  {\bf 82} (2010) 451--497}.

\bibitem{DeFelice2010}
A.~De~Felice and S.~Tsujikawa, \emph{$f(\mathit{R})$ theories},
  \href{http://dx.doi.org/10.12942/lrr-2010-3}{\emph{Living Rev. Relativity}
  {\bf 13} (2010) 3}.

\bibitem{Bertolami2007}
O.~Bertolami, C.~G. B{\"o}hmer, T.~Harko and F.~S.~N. Lobo, \emph{Extra force
  in $f(\mathit{R})$ modified theories of gravity},
  \href{http://dx.doi.org/10.1103/physrevd.75.104016}{\emph{Phys. Rev. D} {\bf
  75} (2007) 104016}.

\bibitem{Harko2008}
T.~Harko, \emph{Modified gravity with arbitrary coupling between matter and
  geometry},
  \href{http://dx.doi.org/10.1016/j.physletb.2008.10.007}{\emph{Phys. Lett. B}
  {\bf 669} (2008) 376--379}.

\bibitem{Harko2010}
T.~Harko and F.~S.~N. Lobo, \emph{$f(\mathit{R},\mathcal{L}_\mathrm{m})$
  gravity}, \href{http://dx.doi.org/10.1140/epjc/s10052-010-1467-3}{\emph{Eur.
  Phys. J. C} {\bf 70} (2010) 373--379}.

\bibitem{Bertolami2010}
O.~Bertolami and J.~P{\'a}ramos, \emph{Mimicking dark matter through a
  non-minimal gravitational coupling with matter},
  \href{http://dx.doi.org/10.1088/1475-7516/2010/03/009}{\emph{J. Cosmol.
  Astropart. Phys.} {\bf 03} (2010) 009}.

\bibitem{Einstein1928}
A.~Einstein, \emph{Riemannian geometry with maintaining the notion of distant
  parallelism}, {\emph{Sitzungsber. Preuss. Akad. Wiss. Phys. Math. Kl.} (1928)
  217}.

\bibitem{Hayashi1979}
K.~Hayashi and T.~Shirafuji, \emph{New general relativity},
  \href{http://dx.doi.org/10.1103/physrevd.19.3524}{\emph{Phys. Rev. D} {\bf
  19} (1979) 3524}.

\bibitem{aldrovandi2012teleparallel}
R.~Aldrovandi and J.~Pereira, \emph{Teleparallel Gravity: An Introduction}.
\newblock Fundamental Theories of Physics. Springer Netherlands, 2012,
  \href{http://dx.doi.org/10.1007/978-94-007-5143-9}{10.1007/978-94-007-5143-9}.

\bibitem{Maluf2013}
J.~W. Maluf, \emph{The teleparallel equivalent of general relativity},
  \href{http://dx.doi.org/10.1002/andp.201200272}{\emph{Ann. Phys. (Berlin)}
  {\bf 525} (2013) 339--357}.

\bibitem{Bengochea2009}
G.~R. Bengochea and R.~Ferraro, \emph{Dark torsion as the cosmic speed-up},
  \href{http://dx.doi.org/10.1103/physrevd.79.124019}{\emph{Phys. Rev. D} {\bf
  79} (2009) 124019}.

\bibitem{Linder2010}
E.~V. Linder, \emph{Einstein's other gravity and the acceleration of the
  universe}, \href{http://dx.doi.org/10.1103/physrevd.81.127301}{\emph{Phys.
  Rev. D} {\bf 81} (2010) 127301}.

\bibitem{Cai2016}
Y.-F. Cai, S.~Capozziello, M.~De~Laurentis and E.~N. Saridakis,
  \emph{$f(\mathit{T})$ teleparallel gravity and cosmology},
  \href{http://dx.doi.org/10.1088/0034-4885/79/10/106901}{\emph{Rep. Prog.
  Phys.} {\bf 79} (2016) 106901}.

\bibitem{Harko2014a}
T.~Harko, F.~S.~N. Lobo, G.~Otalora and E.~N. Saridakis, \emph{Nonminimal
  torsion-matter coupling extension of $f(\mathit{T})$ gravity},
  \href{http://dx.doi.org/10.1103/physrevd.89.124036}{\emph{Phys. Rev. D} {\bf
  89} (2014) 124036}.

\bibitem{Feng2015}
C.-J. Feng, F.-F. Ge, X.-Z. Li, R.-H. Lin and X.-H. Zhai, \emph{Towards
  realistic $f(\mathit{T})$ models with nonminimal torsion-matter coupling
  extension}, \href{http://dx.doi.org/10.1103/physrevd.92.104038}{\emph{Phys.
  Rev. D} {\bf 92} (2015) 104038}.

\bibitem{Li2011}
B.~Li, T.~P. Sotiriou and J.~D. Barrow, \emph{$f(\mathit{T})$ gravity and local
  lorentz invariance},
  \href{http://dx.doi.org/10.1103/physrevd.83.064035}{\emph{Phys. Rev. D} {\bf
  83} (2011) 064035}.

\bibitem{Sotiriou2011}
T.~P. Sotiriou, B.~Li and J.~D. Barrow, \emph{Generalizations of teleparallel
  gravity and local lorentz symmetry},
  \href{http://dx.doi.org/10.1103/physrevd.83.104030}{\emph{Phys. Rev. D} {\bf
  83} (2011) 104030}.

\bibitem{Bahamonde2015}
S.~Bahamonde, C.~G. Böhmer and M.~Wright, \emph{Modified teleparallel theories
  of gravity}, \href{http://dx.doi.org/10.1103/physrevd.92.104042}{\emph{Phys.
  Rev. D} {\bf 92} (2015) 104042}.

\bibitem{Tamanini2012}
N.~Tamanini and C.~G. B{\"o}hmer, \emph{Good and bad tetrads in $f(\mathit{T})$
  gravity}, \href{http://dx.doi.org/10.1103/physrevd.86.044009}{\emph{Phys.
  Rev. D} {\bf 86} (2012) 044009}.

\bibitem{Nunes2016}
R.~C. Nunes, S.~Pan and E.~N. Saridakis, \emph{New observational constraints on
  $f(\mathit{T})$ gravity from cosmic chronometers},
  \href{http://dx.doi.org/10.1088/1475-7516/2016/08/011}{\emph{J. Cosmol.
  Astropart. Phys.} {\bf 08} (2016) 011}.

\bibitem{Maccio2004}
A.~Macci{\`o}, C.~Quercellini, R.~Mainini, L.~Amendola and S.~Bonometto,
  \emph{Coupled dark energy: Parameter constraints from $\mathit{N}$-body
  simulations}, \href{http://dx.doi.org/10.1103/physrevd.69.123516}{\emph{Phys.
  Rev. D} {\bf 69} (2004) 123516}.

\bibitem{Aghanim2009}
N.~Aghanim, A.~C. da~Silva and N.~J. Nunes, \emph{Cluster scaling relations
  from cosmological hydrodynamic simulations in a dark-energy dominated
  universe}, \href{http://dx.doi.org/10.1051/0004-6361/200810692}{\emph{Astron.
  Astrophys.} {\bf 496} (2009) 637--644}.

\bibitem{Baldi2010}
M.~Baldi, V.~Pettorino, G.~Robbers and V.~Springel, \emph{Hydrodynamical
  $\mathit{N}$-body simulations of coupled dark energy cosmologies},
  \href{http://dx.doi.org/10.1111/j.1365-2966.2009.15987.x}{\emph{Mon. Not. R.
  Astron. Soc.} {\bf 403} (2010) 1684--1702}.

\bibitem{Li2011a}
B.~Li, D.~F. Mota and J.~D. Barrow, \emph{$\mathit{N}$-body simulations for
  extended quintessence models},
  \href{http://dx.doi.org/10.1088/0004-637x/728/2/109}{\emph{Astrophys. J.}
  {\bf 728} (2011) 109}.

\bibitem{Gunn1972}
J.~E. Gunn and I.~Gott, J.~Richard, \emph{On the infall of matter into clusters
  of galaxies and some effects on their evolution},
  \href{http://dx.doi.org/10.1086/151605}{\emph{Astrophys. J.} {\bf 176} (1972)
  1}.

\bibitem{Padmanabhan1993}
T.~Padmanabhan, \emph{Structure formation in the universe}.
\newblock Cambridge ; New York : Cambridge University Press, 1993.

\bibitem{Mota2004}
D.~F. Mota and C.~van~de Bruck, \emph{On the spherical collapse model in dark
  energy cosmologies},
  \href{http://dx.doi.org/10.1051/0004-6361:20041090}{\emph{Astron. Astrophys.}
  {\bf 421} (2004) 71--81}.

\bibitem{Maor2005}
I.~Maor and O.~Lahav, \emph{On virialization with dark energy},
  \href{http://dx.doi.org/10.1088/1475-7516/2005/07/003}{\emph{J. Cosmol.
  Astropart. Phys.} {\bf 07} (2005) 003}.

\bibitem{Nunes2006}
N.~J. Nunes and D.~F. Mota, \emph{Structure formation in inhomogeneous dark
  energy models},
  \href{http://dx.doi.org/10.1111/j.1365-2966.2006.10166.x}{\emph{Mon. Not. R.
  Astron. Soc.} {\bf 368} (2006) 751--758}.

\bibitem{Basilakos2009}
S.~Basilakos, J.~C.~B. Sanchez and L.~Perivolaropoulos, \emph{Spherical
  collapse model and cluster formation beyond the $\ensuremath{\Lambda}$
  cosmology: Indications for a clustered dark energy?},
  \href{http://dx.doi.org/10.1103/PhysRevD.80.043530}{\emph{Phys. Rev. D} {\bf
  80} (2009) 043530}.

\bibitem{Basilakos2010}
S.~Basilakos, M.~Plionis and J.~Sol\`a, \emph{Spherical collapse model in time
  varying vacuum cosmologies},
  \href{http://dx.doi.org/10.1103/PhysRevD.82.083512}{\emph{Phys. Rev. D} {\bf
  82} (2010) 083512}.

\bibitem{Brax2010}
P.~Brax, R.~Rosenfeld and D.~Steer, \emph{Spherical collapse in chameleon
  models}, \href{http://dx.doi.org/10.1088/1475-7516/2010/08/033}{\emph{J.
  Cosmol. Astropart. Phys.} {\bf 08} (2010) 033}.

\bibitem{Fernandes2012}
R.~A.~A. Fernandes, J.~P.~M. de~Carvalho, A.~Y. Kamenshchik, U.~Moschella and
  A.~da~Silva, \emph{Spherical ``top-hat'' collapse in
  general-chaplygin-gas-dominated universes},
  \href{http://dx.doi.org/10.1103/PhysRevD.85.083501}{\emph{Phys. Rev. D} {\bf
  85} (2012) 083501}.

\bibitem{Bellini2012}
E.~Bellini, N.~Bartolo and S.~Matarrese, \emph{Spherical collapse in covariant
  galileon theory},
  \href{http://dx.doi.org/10.1088/1475-7516/2012/06/019}{\emph{J. Cosmol.
  Astropart. Phys.} {\bf 06} (2012) 019}.

\bibitem{Li2014}
W.~Li and L.~Xu, \emph{Spherical top-hat collapse of a viscous unified dark
  fluid}, \href{http://dx.doi.org/10.1140/epjc/s10052-014-2870-y}{\emph{Eur.
  Phys. J. C} {\bf 74} (2014) 2870}.

\bibitem{Carames2014}
T.~R.~P. Caram\^es, J.~C. Fabris and H.~E.~S. Velten, \emph{Spherical collapse
  for unified dark matter models},
  \href{http://dx.doi.org/10.1103/PhysRevD.89.083533}{\emph{Phys. Rev. D} {\bf
  89} (2014) 083533}.

\bibitem{Abramo2007}
L.~R. Abramo, R.~C. Batista, L.~Liberato and R.~Rosenfeld, \emph{Structure
  formation in the presence of dark energy perturbations},
  \href{http://dx.doi.org/10.1088/1475-7516/2007/11/012}{\emph{J. Cosmol.
  Astropart. Phys.} {\bf 11} (2007) 012}.

\bibitem{Abramo2009}
L.~R. Abramo, R.~C. Batista, L.~Liberato and R.~Rosenfeld, \emph{Physical
  approximations for the nonlinear evolution of perturbations in inhomogeneous
  dark energy scenarios},
  \href{http://dx.doi.org/10.1103/physrevd.79.023516}{\emph{Phys. Rev. D} {\bf
  79} (2009) 023516}.

\bibitem{Bohmer2008}
C.~G. B{\"o}hmer, T.~Harko and F.~S.~N. Lobo, \emph{The generalized virial
  theorem in $f(\mathit{R})$ gravity},
  \href{http://dx.doi.org/10.1088/1475-7516/2008/03/024}{\emph{J. Cosmol.
  Astropart. Phys.} {\bf 03} (2008) 024}.

\bibitem{Sefiedgar2009}
A.~S. Sefiedgar, K.~Atazadeh and H.~R. Sepangi, \emph{Generalized virial
  theorem in palatini $f(\mathcal{R})$ gravity},
  \href{http://dx.doi.org/10.1103/PhysRevD.80.064010}{\emph{Phys. Rev. D} {\bf
  80} (2009) 064010}.

\bibitem{Capozziello2013}
S.~Capozziello, T.~Harko, T.~S. Koivisto, F.~S. Lobo and G.~J. Olmo, \emph{The
  virial theorem and the dark matter problem in hybrid metric-palatini
  gravity}, \href{http://dx.doi.org/10.1088/1475-7516/2013/07/024}{\emph{J.
  Cosmol. Astropart. Phys.} {\bf 07} (2013) 024}.

\bibitem{Milgrom2014}
M.~Milgrom, \emph{General virial theorem for modified-gravity mond},
  \href{http://dx.doi.org/10.1103/PhysRevD.89.024016}{\emph{Phys. Rev. D} {\bf
  89} (2014) 024016}.

\bibitem{Santos2015}
N.~S. Santos and J.~Santos, \emph{The virial theorem in eddington-born-infeld
  gravity}, \href{http://dx.doi.org/10.1088/1475-7516/2015/12/002}{\emph{J.
  Cosmol. Astropart. Phys.} {\bf 2015} (2015) 002}.

\bibitem{Javadinezhad2016}
R.~Javadinezhad, J.~T. Firouzjaee and R.~Mansouri, \emph{Relativistic virial
  relation for cosmological structures},
  \href{http://dx.doi.org/10.1103/physrevd.93.023007}{\emph{Phys. Rev. D} {\bf
  93} (2016) 023007}.

\bibitem{Navarro1997}
J.~F. Navarro, C.~S. Frenk and S.~D.~M. White, \emph{A universal density
  profile from hierarchical clustering},
  \href{http://dx.doi.org/10.1086/304888}{\emph{Astrophys. J.} {\bf 490} (1997)
  493}.

\bibitem{Weitzenbock1923}
R.~Weitzenb{\"o}ck, \emph{Invarianten-Theorie}.
\newblock P. Noordhoff, 1923.

\bibitem{Gron2007}
{\O}.~Gr{\o}n and S.~Hervik, \emph{Einstein's General Theory of Relativity:
  With Modern Applications in Cosmology}.
\newblock Springer New York, 2007,
  \href{http://dx.doi.org/10.1007/978-0-387-69200-5}{10.1007/978-0-387-69200-5}.

\bibitem{Bertolami2008}
O.~Bertolami, F.~S.~N. Lobo and J.~P{\'a}ramos, \emph{Nonminimal coupling of
  perfect fluids to curvature},
  \href{http://dx.doi.org/10.1103/physrevd.78.064036}{\emph{Phys. Rev. D} {\bf
  78} (2008) 064036}.

\bibitem{Minazzoli2012}
O.~Minazzoli and T.~Harko, \emph{New derivation of the lagrangian of a perfect
  fluid with a barotropic equation of state},
  \href{http://dx.doi.org/10.1103/physrevd.86.087502}{\emph{Phys. Rev. D} {\bf
  86} (2012) 087502}.

\bibitem{Harko2014}
T.~Harko, F.~S. Lobo, G.~Otalora and E.~N. Saridakis, \emph{$f (\mathit{T}
  ,\mathcal{T})$ gravity and cosmology},
  \href{http://dx.doi.org/10.1088/1475-7516/2014/12/021}{\emph{J. Cosmol.
  Astropart. Phys.} {\bf 12} (2014) 021}.

\bibitem{Passare2004}
M.~Passare and A.~Tsikh, \emph{The Legacy of Niels Henrik Abel}, ch.~Algebraic
  Equations and Hypergeometric Series, pp.~653--672.
\newblock Springer Science, 2004.

\bibitem{Jackson1970}
J.~C. Jackson, \emph{The dynamics of clusters of galaxies in universes with
  non-zero cosmological constant, and the virial theorem mass discrepancy},
  \href{http://dx.doi.org/10.1093/mnras/148.3.249}{\emph{Mon. Not. R. Astron.
  Soc.} {\bf 148} (1970) 249--260}.

\bibitem{Debbasch2009}
F.~Debbasch and W.~van Leeuwen, \emph{General relativistic boltzmann equation,
  $\mathit{II}$: Manifestly covariant treatment},
  \href{http://dx.doi.org/10.1016/j.physa.2009.01.009}{\emph{Physica A} {\bf
  388} (2009) 1818--1834}.

\bibitem{MILNE1934}
E.~A. Milne, \emph{A newtonian expanding universe},
  \href{http://dx.doi.org/10.1093/qmath/os-5.1.64}{\emph{Q. J. Math.} {\bf
  os-5} (1934) 64--72}.

\bibitem{Moore1994}
B.~Moore, \emph{Evidence against dissipation-less dark matter from observations
  of galaxy haloes}, \href{http://dx.doi.org/10.1038/370629a0}{\emph{Nature}
  {\bf 370} (1994) 629--631}.

\bibitem{Boylan-Kolchin2012}
M.~Boylan-Kolchin, J.~S. Bullock and M.~Kaplinghat, \emph{The milky way's
  bright satellites as an apparent failure of $\lambda$cdm},
  \href{http://dx.doi.org/10.1111/j.1365-2966.2012.20695.x}{\emph{Mon. Not. R.
  Astron. Soc.} {\bf 422} (2012) 1203--1218}.

\end{thebibliography}\endgroup
\bibliographystyle{JHEP}
\appendix
\section{Roots of the principal quintic}
\label{appendix}
As is known to all,
a quintic equation does not have a
general expression of roots.
However, for principal quintic
\begin{equation}
	m x^5+n x^2+x+1=0,
	\label{pquintica}
\end{equation}
Passare and Tsikh\cite{Passare2004} found that
a root can be expressed as a series expansion:
\begin{equation}
	x_5=-\sum_{j,k\ge0}\frac{\left( 2j+5k \right)!}{j!k!\left( j+4k+1 \right)!}n^j\left( -m \right)^k,
	\label{seriesroot}
\end{equation}
with the domain of convergence given by the condition
\begin{equation}
	\begin{split}
	&3125|m|^2-4^4|m|+2^23^3|n|^5-3^3|n|^4+25^2|m||n|\\
	&-2\cdot3^25^3|n|^2|m|<0.
\end{split}
	\label{conditiona}
\end{equation}

Eq.\eqref{quintic} can be written in the form of Eq.\eqref{pquintica}
with
\begin{equation}
	x=\frac{\tilde k\beta}{\Omega_\text{mi}(1+\delta_\text{i})},
	\label{redef1}
\end{equation}
and
\begin{equation}
	m=-\frac{\tilde B\Omega_\text{mi}^4(1+\delta_\text{i})^4}{k^6},\quad n=-\frac{\tilde A\Omega_\text{mi}(1+\delta_\text{i})}{k^3}.
	\label{redef2}
\end{equation}
For $\tilde A,\tilde B>0$ and $\tilde k<0$,
we have $m<0$ and $n>0$.
And then Eq.\eqref{seriesroot}, if converges,
gives a negative real root $x_5<0$.
Since $\tilde k<0$,
Eq.\eqref{redef1} then gives
a corresponding positive real root of Eq.\eqref{quintic}
denoted as $\beta_5>0$.
Thus, Eq.\eqref{conditiona} with Eq.\eqref{redef2}
gives a sufficient condition for Eq.\eqref{quintic}
to have real positive root(s).

As for the necessity of Eq.\eqref{condition},
we know that Eq.\eqref{pquintica} will have
at least one real root, denoted as $x_1$.
And $x_1>0$ because $m<0$.
Since the derivative of Eq.\eqref{pquintica},
$5mx^4+2nx+1=0$,
only has two real roots,
the quintic function
$mx^5+nx^2+x+1$
has only two extrema.
So the principal quintic equation \eqref{pquintica}
has at most three real roots.
That is, Eq.\eqref{pquintica}
has a real root
and a pair of conjugate complex roots for sure.
This guaranteed pair of conjugate complex roots are denoted as $x_2,x_3$.
Now $x_5$ is either real (when Eq.\eqref{seriesroot} converges)
or complex (when Eq.\eqref{seriesroot} diverges)
depending on $m,n$,
therefore it will not be one of the three roots $x_1,x_2,x_3$.
Assume that the condition Eq.\eqref{conditiona} is falsified
and hence Eq.\eqref{seriesroot} diverges and $x_5$ is complex.
According to Vieta's relation among the roots,
the last root $x_4=-(x_1+x_2+x_3+x_5)$.
Since $x_2,x_3$ are conjugate complex roots and $x_5$ is complex,
$x_4$ is complex, too.
In other word,
if Eq.\eqref{conditiona} is falsified,
Eq.\eqref{pquintica} will have only one real root,
and it is positive,
and hence Eq.\eqref{quintic} will not have any real positive root.
Thus, Eq.\eqref{condition} is the necessary and sufficient
condition for Eq.\eqref{quintic} to have real positive root(s).
\end{document}